\documentclass[a4paper, 11pt]{article}    
\usepackage{latexsym}
\usepackage{amssymb}
\usepackage{amsmath}
\usepackage{amsfonts}
\usepackage{bbm}
\usepackage{graphicx}
\usepackage{float}
\usepackage[english]{babel}
\usepackage{multirow}
\usepackage{float}
\usepackage{soul,color}
\restylefloat{table}
\usepackage{caption}
\usepackage{placeins}
\usepackage{hyperref}
\usepackage{wrapfig}
\usepackage{cleveref}
\usepackage{enumitem}
\usepackage[toc,page]{appendix}
\usepackage[normalem]{ulem}
\useunder{\uline}{\ul}{}

\makeatletter         
\def\@maketitle{
\begin{center}
{\Huge \bfseries \sffamily \@title }\\[4ex] 
{\Large  \@author}\\[2ex] 
\includegraphics[width = 30mm]{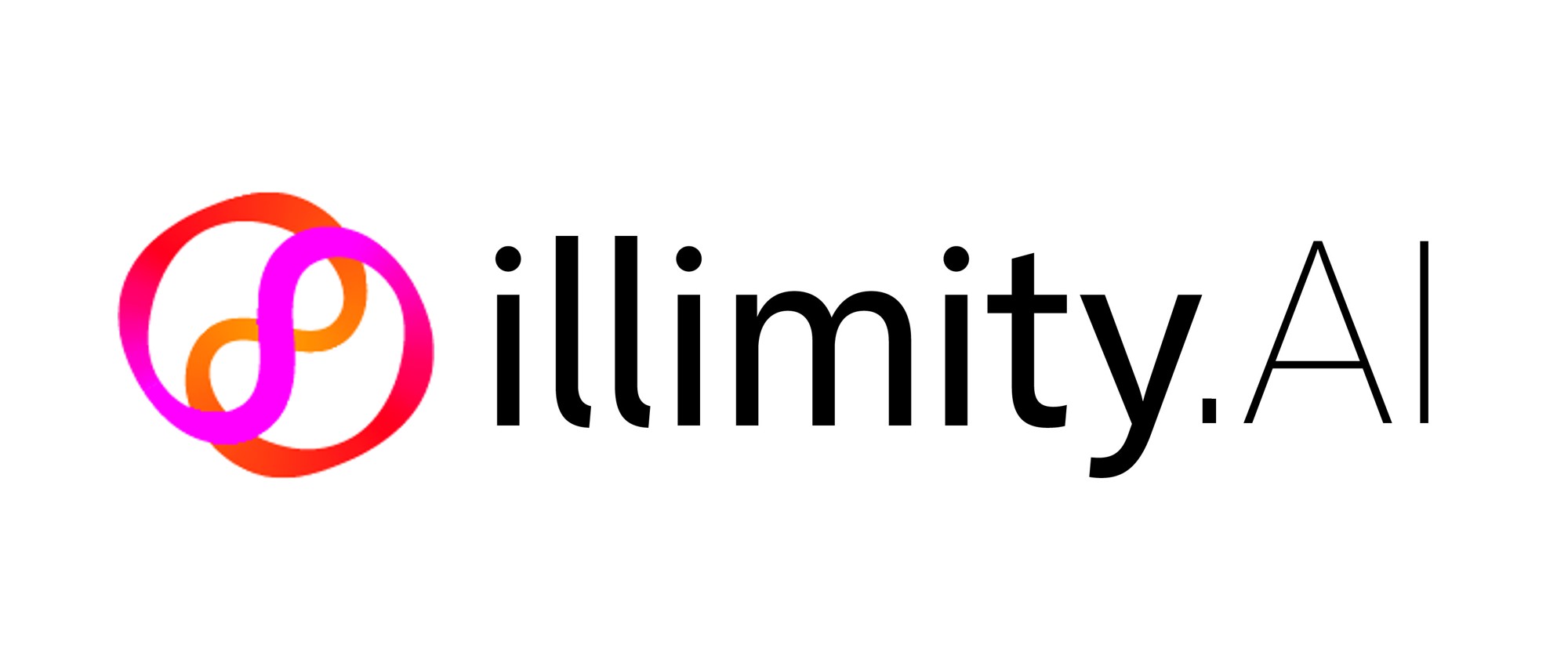}\\[4ex]
\@date\\[4ex]
\end{center}}
\makeatother
\title{ \bf An Artificial Intelligence approach to Shadow Rating}  
\author{A. R. Provenzano\footnote{corresponding author: angela.provenzano@illimity.com}\,\,, D. Trifirò, N. Jean, G. Le Pera, M. Spadaccino, L. Massaron and C. Nordio}
\begin{document}           
\maketitle                 

\begin{center}
\tt \Large Working paper\footnote{This paper reflects the authors' opinions and not necessarily those of their employers.}
\end{center}

\vspace{10mm}

\begin{abstract} 
\it \noindent

We analyse the effectiveness of modern deep
learning techniques in predicting credit ratings over a universe of thousands of global corporate
entities obligations when compared to most popular, traditional machine-learning approaches
such as linear models and tree-based classifiers. Our results show a adequate accuracy over
different rating classes when applying categorical embeddings to artificial neural networks
(ANN) architectures.

\end{abstract}

\bigskip
{\bf JEL} Classification codes: C45, C55, G24
{\bf AMS} Classification codes: 62M45, 91G40
\bigskip



\section*{Introduction}
\label{section:intro}

Credit ratings are alphabetical indicators of credit risk provided by international rating agencies, such as Standard \& Poors (S\&P), Moody's and Fitch, used for ever increasing applications in both financial markets and banking risk management \cite{mcneil2005quantitative}. Rating agencies claim to use both quantitative and qualitative information in arriving at a rating, but do not reveal the methodological detail of rating assignment. The process of mimicking external ratings when there is insufficient data to build an explicit default prediction model is known as \emph{Shadow Rating Approach} (SRA)\cite{erlenmaier2006shadow}. \\
Using financial and business data to assign a credit rating is a challenging process due to the complex and non-linear interactions between variables that are supposed to be good predictors of future company defaults, such as balance-sheet factors or ratios, macroeconomic indicators, qualitative information about the company (such as the company's competitive arena). Moreover, rating assignment process lacks a well-defined theory, and this makes it difficult to apply conventional mathematical or rule based techniques. For all these reasons, numerous quantitative approaches have been developed\cite{toseafa2019predicting}. Among all of them, Artificial Neural Networks (ANN) demonstrate to have particular strength in classifying outcomes, as they are able to derive meaningful information from data without having to make assumptions with respect to the underlying properties and relationships within the input variables \cite{heaton2016empirical}, nor to rely on explicit knowledge of parameters or prior specification of theoretical models. Indeed, a successful Neural Network implementation create a system of relationships that, learning from past examples, is able to generalize these "lessons" to new ones. Dutta \& Shekhar (1988) were the first to develop an ANN model for corporate bond rating \cite{dutta1988bond}. They compared the performance of ANN models with regression models and found that ANN consistently performed better in predicting bond ratings from a given set of financial ratios, motivating further implementations of ANN for bond ratings. Bennell et al. \cite{bennell2006modelling} show that ANN represent a superior technology for calibrating and predicting sovereign rating with respect to \emph{Ordered Probit Modelling}, which has been considered by the previous literature to be the most successful econometric approach.\\Whilst the search for increasing model accuracy has encouraged the development of more and more complex models, the ability to interpret outputs of a model has become as crucial as the prediction's accuracy in many applications, such as medical research, or in strongly regulated fields, such as the banking sector, in which simple models (like linear or logistic regressions) are often preferred for their ease of interpretation. \\In this article we investigate the modelling and predictive performance of neural networks reinforced with \emph{embeddings}, against linear regression, logistic regression and simple neural networks algorithms for a sample of European corporate ratings provided by Moody's. Then we provide an intuitive way to interpret the model predictions, in order to prevent model bias, encourage the use of machine learning model in decision making process, engender trust in model predictions and provide insight for model improvement. In more details, the article is organized as follows: \Cref{section:data} is dedicated to describe the dataset of our analysis; in \Cref{section:model} the model architecture is presented, together with a rapid description of a selection of alternative machine learning models used as baseline for our approach; models results and are showed in \Cref{section:results}.


\section{Data}
\label{section:data}

\begin{figure}[h!]
    \centering
    \includegraphics[scale=0.6]{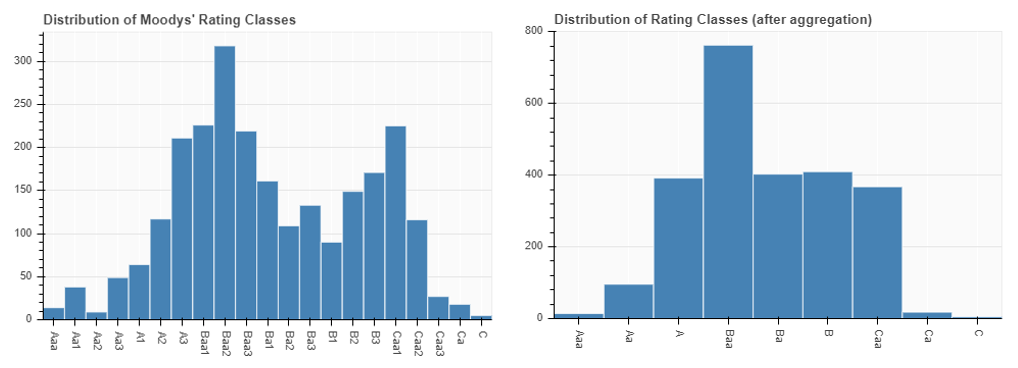}
    \caption{Histogram of the rating classes: on the left the original classification provided by Moody's, on the right the aggregated classes}
    \label{fig:histo_rating}
\end{figure}

To train our model and evaluate its performance we collected a sample of 2469 annual (end of calendar year) observations of corporate credit ratings assigned by Moody's to 509 companies across Europe\footnote{Austria, Belgium, Denmark, France, Germany, Greece, Ireland, Italy, Luxembourg, Netherlands, Portugal, Slovenia and Spain}, during the period from 1996 to 2018. Input variables of the model are represented by balance-sheet indexes and ratios, and \emph{Key Performance Indicators} (KPI) calculated from Orbis's financial reports\cite{orbis}. The input features were selected to be consistent with factors that can affect the capacity and willingness of borrowers to service external debt and included indicators for efficiency, liquidity, solvency and profit. Our analyses have been focusing exclusively on corporate debt, thus excluding both financial institutions and sovereign issues, covering different sectors (e.g. aerospace, automotive, construction, real estate, consumer goods durable and not-durable, energy, high-teach industries, media, etc.). Output variables were extracted from \emph{Moody’s Default $\&$ Recovery Database} (DRD)\cite{DRD} which provides information about rated entities, rated defaulters, and unrated defaulters back to 1980 for EMEA from several sectors (e.g. corporates, sovereigns, sub-Sovereigns, financial institutions, insurance companies). Target variable has been defined as Moody’s Senior Rating Standard, i.e. Issuer-level rating, Senior Unsecured or Equivalent, based on the full debt structure of the company. On specifics, Moody's rating scale comprises 21 notches\footnote{One rating notch is defined as the difference between two consecutive rating classes}, running from a high of Aaa to a low of C, divided into two sections: \emph{investment grade} (i.e. higher than Baa3) and \emph{speculative grade} (lower than Ba1) (see \Cref{table:rating_scale}). Because of a highly imbalanced dataset between the rating classes (as shown in the left histogram of \Cref{fig:histo_rating}), we have simplified the classification problem by aggregating the original 21 notches into 9 wider classes (see \Cref{table:rating_scale}).

\begin{table}[h!]
\centering

\begin{tabular}{|c|c|c|}
\hline
\multicolumn{2}{|c|}{\textbf{Moody’s rating scale}}            & \textbf{Aggregated rating scale} \\ \hline
\multirow{4}{*}{\textit{Investment grade}} & Aaa    & Aaa                              
\\
                                            & Aa1, Aa2, Aa3      & Aa                          
\\
                                            & A1, A2, A3       & A                                \\
                                            & Baa1, Baa2, Baa3 & Baa                              \\ \hline
\multirow{5}{*}{\textit{Speculative grade}} & Ba1, Ba2, Ba3    & Ba                               \\ 
                                            & B1, B2, B3       & B                                \\ 
                                            & Caa1, Caa2, Caa3 & Caa
\\                         
                                            & Ca               &                              Ca
                                            \\
                                            & C                &                                 C \\ \hline

\end{tabular}
\caption{Moody's rating scale (on the left) and rating classes after aggregation (on the right)}
\label{table:rating_scale}

\end{table}

In order to allow the model to perform more accurately, we joined original information with more general macro variables addressing the surrounding climate in which companies operate.
Among the wide range of macroeconomic indicators provided by Oxford Economics \cite{oxford}, a subset of the most influential ones has been selected as explanatory variables. Some of them are country-specific, others are common to the whole Eurozone\footnote{Regional aggregate \emph{Eurozone} includes the following countries: Austria, Belgium, Cyprus, Estonia, Finland, France, Germany, Greece, Ireland, Italy, Latvia, Lithuania, Luxembourg, Malta, Netherlands, Portugal, Slovenia, Slovakia and  Spain}.

\paragraph{Country-specific indicators}
	\begin{enumerate}[itemsep=0.0mm]
		\item \emph{Interest rate, central bank policy}: the rate that is used by central bank to implement or signal its monetary policy stance (expressed as an average).
		\item \emph{Government debt, Maastricht definition}: the total outstanding borrowing of central and local governments expressed in local currency according to the convergence criteria set out in the Maastricht Treaty.
		\item \emph{Credit rating}: the sovereign risk rating that varies between 0 (default) and 20 (AAA rating). It is based on the average of the sovereign ratings provided by Moody’s, S\&P and Fitch.
		\item \emph{GDP}: The volume of all final goods and services produced within a country in a given period of time in local currency terms.
		\item \emph{Interest rate, 10-year government bond yield}: government bonds maturing in ten years.
		\item \emph{Interest rate, 5-year government bond yield}: government bonds maturing in five years.
		\item \emph{Interest rate, short-term, 3 month t-bill rate}: 3 month treasury-bill rate.
		\item \emph{Share price index}: index of most liquid shares listed in national equity stock-market.
		\item \emph{Stock-market dividends index}: stock-market dividends index calculated as: stock-market dividend yield $\times$ Share price index, Global Equity Indices / 100.
	\end{enumerate}

\paragraph{Eurozone indicators}
	\begin{enumerate}[itemsep=0.0mm]
		\item \emph{Interest rate, 6-month}: the 6-month interbank rate.
		\item \emph{Interest Rate Swap, 2-year}: swap par-rate for 2 years tenor.
		\item \emph{Interest Rate Swap, 5-year}: swap par-rate for 5 years tenor.
		\item \emph{Interest Rate Swap, 10-year}: swap par-rate for 10 years tenor.
		\item \emph{Interest Rate Swap, 30-year}: swap par-rate for 30 years tenor.
		\item \emph{Interest rate, EONIA}: 1-day interbank interest rate for the Eurozone.
		
	\end{enumerate}

\section{Model}
\label{section:model}

The task of this study consists in replicating the ratings that Moody's assigns to each company, given its balance-sheet conditions and macro-economic factors. It is important to point out that our attempt is to replicate the rating assigned by a human agent rather than providing an independent, specific rule-based approach. \\The approach we propose in this article consists of an \emph{Artificial Neural Network} model with \emph{Categorical Embedding} (ANN EMB model) in which the output layer is a continuous value between 0 and 8, corresponding respectively to the highest and the lowest rating classes (i.e. Aaa - C). Thus this can be considered a classical supervised learning problem, where provided labels (rating classes in our case) are used as targets for algorithms to learn linear and non linear relationships with selected independent variables. In the Machine Learning jargon a problem in which you are interested in predicting the value of a dependent variable (i.e. the Target) via modeling the relationship between it and one or more independent variables (the features input the model) is know as a \emph{regression problem}. The predicted value obtained via ANN EMB model is then mapped into a discrete rating class as reported in \Cref{table:rating_map}.

\begin{table}[h!]
\centering

\begin{tabular}{|c|c|}
\hline
\textbf{Rating class} &\textbf{Target value}\\
\hline
Aaa    & 0 \\
Aa & 1 \\
A & 2 \\
Baa & 3 \\
Ba & 4 \\
B & 5\\
Caa & 6 \\
Ca & 7\\
C & 8\\
\hline

\end{tabular}
\caption{Rating classes (on the left) and corresponding target values (on the right)}
\label{table:rating_map}

\end{table}

\subsection{Neural Networks}

\begin{figure}[h!]
     \centering
	\includegraphics[scale=0.45]{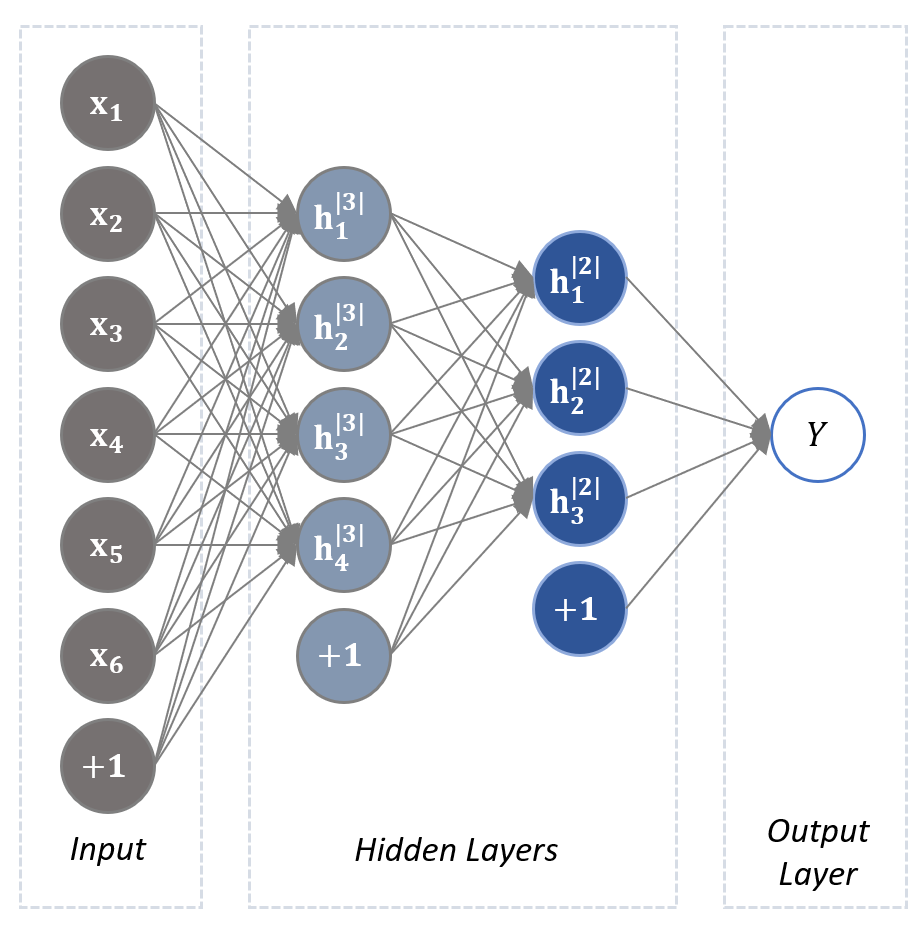}
     \caption{Scheme of continuous output Neural Network}
     \label{fig:perceptron}
\end{figure}

Among the most well known and used approaches \emph{Neural Networks} are probably the most widely popular when dealing with complex and non-linear problems. The \emph{Perceptron}\cite{geron2017hands} is one of the simplest \emph{Artificial Neural Networks} architectures composed of a single layer of \emph{Linear Threshold Units} (LTUs), artificial neurons in which inputs and outputs are numbers, each input connection is associated with a weight and each neuron is connected to all the inputs. The sum product of the weights and input signals are processed by its activation function to produce an output signal. In case of a deep neural network, more layers of these densely connected neurons are added sequentially, allowing for further complexity and non-linearity of the model. In case of continuous output, a final neuron is added to the model architecture that allows for a linear activation of all the outputs coming from the previous layers (a scheme is provided in \Cref{fig:perceptron}). 

\subsection{Feature engineering and Categorical Embeddings}

Before using data in each machine learning process, one important task consists in generating a viable set of features that facilitate convergence towards the desired objective. This task normally requires data manipulation, e.g. transforming categorical features, treating missing values, replacing outliers. \\Categorical, non-ordinal features are one of the main issues that must be tackled before using any models, since they are often not available as numbers (which are the only processed data that can feed any machine learning algorithm). Although a simple numerical transformation (i.e. \emph{Label Encoding}) can be sufficient in some approaches (mainly, \emph{Tree-based techniques}), further attention needs to be paid in case of linear models like \emph{Logistic Regressions} and Neural Networks, since the way they are built would imply unnecessary, non-existent ordinal relationships that could fool the model during the training process. One normal way to address the problem of transforming categorical, non-numerical variables is to map it into a list of random integers; a further step which is needed for architectures such as Neural Networks is also to transform this mapping into a fixed size sparse vector that consists of all zeros but 1 in the cell used uniquely to identify specific realization of that variable (i.e. \emph{One-Hot Encoding}). Although this approach is widely used in Neural Networks and linear models (where simple encoding of non-ordinal categorical variables would make it hard for the model to extract the right relationship between target and the specific realizations of those variables) the main drawbacks rely on the fact that categories with high-cardinality (i.e. an high number of possible realizations of that category) would generate large dimension datasets that, apart from memory issues, would not help the process of loss minimization of the algorithm (the well known "\emph{curse of dimensionality}"). Furthermore, there is no way to represent the similarity between words, i.e. One-Hot encoding does not place similar entities closer to one another in vector space.

An alternative more recent approach, that overcomes the problems of the One-Hot encoding technique is known as \emph{Categorical Embedding}. It consists in mapping each possible realization of a given categorical feature in a multi-dimensional vector, with weights that are directly learnt by the neural network during the process of training via \emph{back-propagation}. In the context of neural networks, embeddings are then low-dimensional, learned continuous vector representations of discrete variables. This technique, together with a significant speeding up of the learning process and reduction in data dimensionality and consequent memory usage, allows placing each category in a Euclidean space keeping coherent relationship with other realization of the same feature. This property has been effectively been capitalized.

An example of categorical embedding is \emph{Word Embedding}\cite{guo2016entity}: it is a class of approaches for representing words and documents using a dense vector representation, where a vector represents the projection of the word into a continuous vector space. It is an improvement over the more traditional \emph{bag-of-word} encoding schemes, where large sparse vectors are used to score each word within a vector to represent an entire vocabulary.

\subsection{Model architecture and data processing}

The ANN architecture proposed in this article have been prototyped via Keras  \footnote{Keras is a very popular, high-level deep learning framework, written in Python by François Chollet.}\cite{gulli2017deep} and consists of 2 layers, 64 neurons per layer, and requires 150 epochs to converge. In order to control the magnitude of change in the network weight between one iteration and the next, a learning rate of 0.01 is applied. To prevent \emph{overfitting} the Train set during the Cross Validation phase (which will be widely described hereafter), the \emph{early stopping} technique has been used. It consists of interrupting the training phase when performance on the validation set starts dropping: in fact, as validation accuracy captures the generalization capabilities of the model, in case of overfitting it stops increasing and can even start to decrease. The metric used for early stopping purpose is the \emph{Quadratic Weighted Kappa Function} that will be discussed in \Cref{section:results}. As a further regularization technique \emph{dropout}\cite{srivastava2014dropout} has been performed. This technique consists in randomly drop-out units (hidden and visible) along with their incoming and outgoing connections from the neural network during training, but potentially activating it during the next step. The choice of which units to drop is random: in the simplest case, each unit is retained with a fixed probability $p$\footnote{The so called \emph{dropout-rate} $p$ is a tunable hyper-parameter that can simply be set at $0.5$, which seems to be close to optimal for a wide range of networks and tasks} independent of other units.\\ \emph{K-fold Cross-Validation} technique has been used both for \emph{hyper-parameters tuning}, i.e. to configure model-specific properties such as its complexity (e.g. number of layers, number of neurons per layer, number of epochs) or how fast it should learn from data (e.g. learning rate), and for evaluating accuracy metrics of the model to the whole dataset. It consists in dividing the dataset into K subsets, training the model on K-1 subsets and holding the last one for testing the model performance, repeating the process K times. The error estimation is averaged over all K trials to get total effectiveness of the model. In order to ensure that each fold contains approximately the same percentage of samples of each classes as the complete set \emph{Stratified K-Fold} has being used. This variant of Cross Validation technique makes sure that each fold is a good representative of the whole dataset. \\For each fold, \emph{imputation} of the raw data has been performed. In particular, for each pair of train/validation-set missing values have been replaced with the median of the non-missing values calculated within each column of the train-set separately and independently from the others. Then categorical features have been transformed via One-Hot Encoding, except for the \emph{Nace2 Description}\footnote{Nace2 is a statistical classification of economic activities valid throughout the whole European Community Countries and subject of legislation at a European Union level} processed via \emph{Word Embedding}. In order to improve model performance, numerical data must have the same scale, so for these reasons continuous-valued features have been transformed via \emph{quantile-normalization}\footnote{This technique consist of transforming each feature to follow a uniform or normal distribution. First an estimate of the cumulative distribution function of a feature is used to map the original values to a uniform distribution. The obtained values are then mapped to the desired output distribution using the associated quantile function. Features values of new/unseen data that fall below or above the fitted range will be mapped to the bounds of the output distribution.} and \emph{scaling} by fitting the data on the k-th train-set and than transforming the k-th pair of train/validation-set for each fold.  

\begin{figure}[h!]
     \centering
     \vspace{-10pt}
	\includegraphics[scale=0.37]{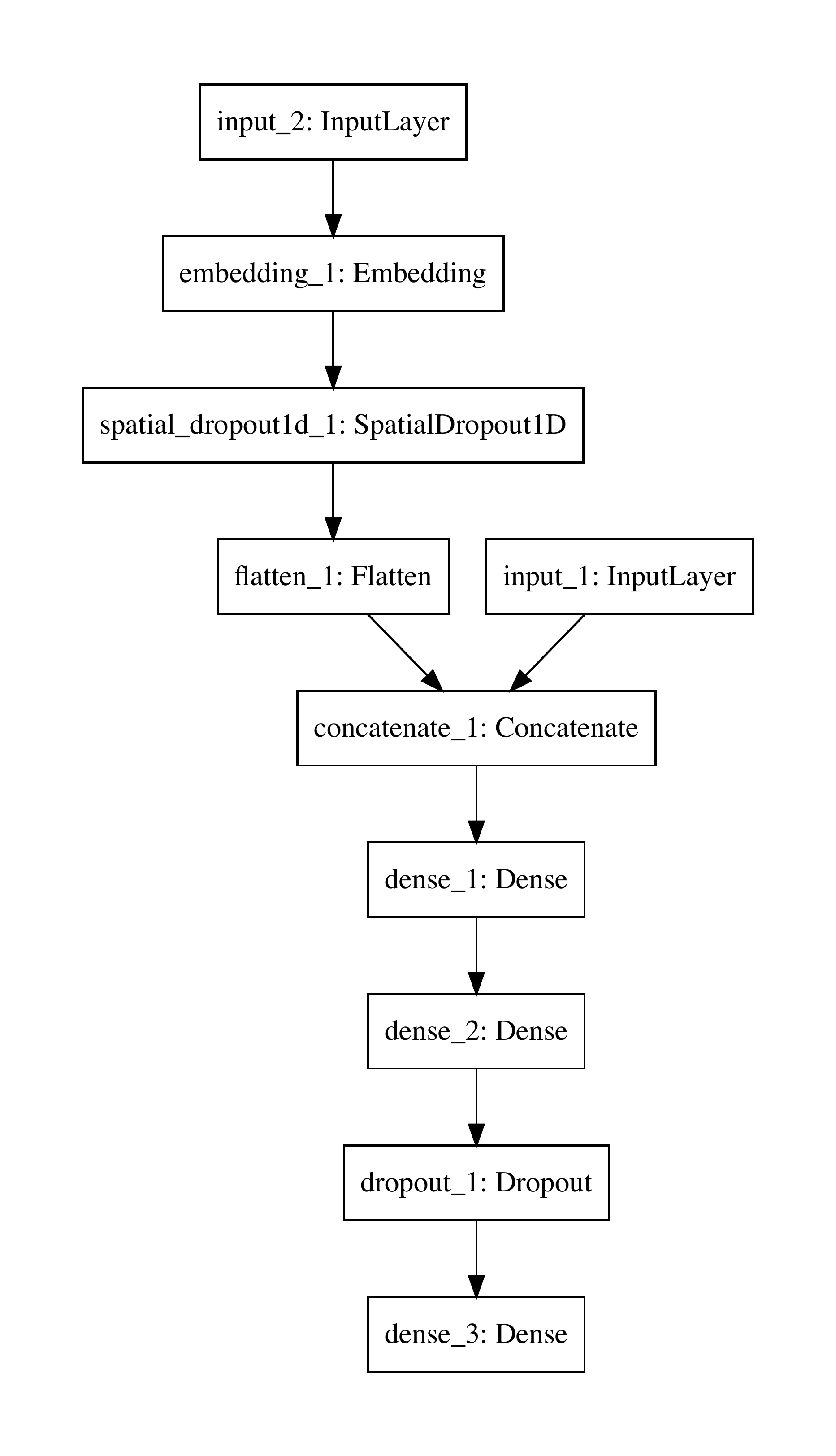}
	\vspace{-10pt}
     \caption{Scheme of Keras ANN with Embedding architecture}
     \label{fig:ANN_EMB_architecture}
\end{figure}

Neural Network architecture is reported in \Cref{fig:ANN_EMB_architecture}. 

At first, feature "\emph{Nace2 Description}" has been embedded: the sequence of \emph{Nace2} sector textual field was fed into an array of embedding layers of a given size (in our case, the number of words in the longest sentence of the "Nace2 Description"). The embedding layer has been initialized by default to random values since no pre-trained embeddings are used. Categorical embeddings obtained so far have then been regularized via \emph{Spatial 1D} version of dropout \footnote{It allows to drop entire 1D feature maps instead of individual elements, and it is quite useful in presence of strong correlation between feature maps, as is normally the case in early convolution layers}. Finally, those embeddings have been concatenated to both continuous numeric features and one-hot encoded categorical features, and passed as the final input to a dense layer.

\subsection{Alternative models}

In order to provide a baseline to our approach, we selected a bunch of alternative machine learning algorithms that have been widely used in the field of credit rating prediction. On the whole list of this models hyper-parameters optimization have been executed via stratified cross-validation, in order to maintain an adequate proportion of all rating classes in all the folds used in the process.

\paragraph{Linear Regression}

The simpler approach consists in applying a linear regression by computing a weighted sum of all the features (plus a bias term) with weights optimised in order to minimize a loss measure (in this case the \emph{Root Mean Square Error} ($RMSE$)) between the predicted rating output and the ground truth. The $RMSE$ gives the idea of how much error the system typically makes in its predictions, and is defined as follows:

\begin{equation}
RMSE(\mathbf{X}, h) = \sqrt{\frac{1}{m} \sum_{i=1}^m \left(h(\mathbf{x}^{(i)}) - y^{(i)} \right)^2}
\end{equation}

where $m$ is the number of instances in the dataset you are measuring the $RMSE$ on, $\mathbf{x}^{(i)}$ is a vector of all the feature values of the $i^{th}$ instance in the dataset and $y^{(i)}$ is the desired output value for that instance and $h$ is the system's predictions function (\emph{hypothesis}). This approach by definition cannot capture any non-linear relationship between features and target or interaction between features.

\paragraph{Logistic Regression}

Since regression models represent a standard approach for credit risk analyses (e.g. bond rating, bankruptcy prediction, etc.) \emph{Logistic Regression} is a benchmark by which ANN implementations have to be compared. A Logistic Regression (LR) model is commonly used as a binary classifier to estimate the probability that an instance $x$ belongs to a particular class. Just like a linear regression model, it computes a weighted sum of the input features (plus a bias term), but instead of outputting the result itself, it calculates its logistic\footnote{The \emph{logistic} is a sigmoid function $ y(x)=\frac{1}{1+e^{-x}}$ that outputs a number between 0 and 1.} transformation. Logistic Regression model can be easily generalized to the case of multilabel classification problem via \emph{one-vs-rest} (also known as \emph{one-vs-all}) technique. It consists in training $C$ separate binary classification models, where each classifier $f_c$ is trained to determine whether or not an example is part of class $c \in \{c_0, c_1, \dots, C\}$ or not. To predict the class for a new example $x$, we run all classifiers on it and choose the class with the highest score: 

\begin{equation} 
\label{eqn:logistic}
\widehat{y} = argmax_{c \in \{c_0, c_1, \dots, C\}}\{f_c(x)\}
\end{equation}

One main drawback of this technique is that when the number of classes is high, each binary classifier has to deal with a highly imbalanced dataset, which may deteriorate the whole model performance.

\paragraph{Artificial Neural network without Embeddings}

In order to show how embedding of \emph{Nace2 Description} sectors is helpful in improving overall performance of our model, a simpler version of the ANN has been tested, allowing only the traditional one-hot encoding of all categorical features (thus including also our \emph{Nace2 Description} sector feature).

\section{Results}
\label{section:results}

The metric used to evaluate the model performance is the \emph{Quadratic Weighted Kappa Metric} (qwk)\cite{brenner1996dependence}, which coincides with the \emph{Cohen's kappa}, defined in \Cref{eqn:k}, with quadratic weights\footnote{in which the magnitude of the discrepancy between ratings are proportional to the square of the deviation of individual ratings}. In particular, the Cohen's kappa is calculated between the scores which are expected/known and the predicted scores as follows: 

\begin{equation}\label{eqn:k}
k = \frac{p_{o}-p_{e}}{1-p_{e}}
\end{equation}

where $p_{o}$ is the empirical probability of agreement on the label assigned to any sample (the observed agreement ratio), and $p_{e}$ is the expected agreement when both annotators assign labels randomly. The qwk is a score that measures the level of agreement between two ratings on a classification problem and it varies between $-1.0$ and $1.0$: a coefficient of $1.0$ indicates maximum possible agreement between raters, a coefficient of $0.0$ indicates random agreement between raters, while negative values may occur in the case of less than chance agreement. \\Prediction results are summarized into a \emph{confusion matrix} which counts the number of correct and incorrect predictions for each rating class. In details, given our $N=9$ rating classes, the confusion matrix would be an $N \times N$ matrix, with the left axis showing the true class and the top axis showing the class assigned to an item with that true class. Each element $i$,$j$ of the matrix would be the number of items with true class $i$ that were classified as being in class $j$. \\More concise metrics are used to compare our model to the alternative machine learning algorithms described above: the \emph{Precision} (see \cref{eqn:precision}), i.e. the accuracy of the positive predictions, and the \emph{Recall} (see \cref{eqn:recall}), i.e. the true positive rate. In particular, the Precision is defined as follows:

\begin{equation}\label{eqn:precision}
Precision = \frac{TP}{TP+FP}
\end{equation}

where $TP$ is the number of true positives, and $FP$ is the number of false positives. While the Recall is the ratio of positive instances that are correctly detected by the classifier:

\begin{equation}\label{eqn:recall}
Recall = \frac{TP}{TP+FN}
\end{equation}

where $FN$ is the number of false negatives. Precision and Recall are combined into a single metric called \emph{F$_{1}$-score}, quite useful to compare two or more classifiers. It is defined as the harmonic mean of the aforementioned metrics (see \cref{eqn:f1_score}):

\begin{equation}\label{eqn:f1_score}
F_1 = \frac{Precision \times Recall}{Precision + Recall} = \frac{TP}{TP + \frac{FN + FP}{2}}
\end{equation}

\subsection{ANN EMB Model performance}

Considering the complexity of the problem of rating prediction, we obtained a satisfying performance of the model in term of Quadratic Weighted Kappa Metric: $0.864 \pm 0.010$. Only for classes with very few observations, the model shows some weakness in predicting the correct results, i.e. classes Aaa (target value 0), and C (target value 8). From our analysis, we can see that our ANN EMB over-performs the other models. Details of Precision, Recall and F$_{1}$-score metrics are reported in \Cref{table:ANN_EMB_Results}.

\begin{table}[h!]
\centering

\begin{tabular}{c|r|r|r|r|r|r|r|r|r|}
\cline{2-10}
\textbf{}                                & \multicolumn{9}{c|}{\textit{Rating Classes}}                                                                                                                                                                             \\ \cline{2-10} 
\textit{}                                & \multicolumn{1}{c|}{\textbf{Aaa}} & \multicolumn{1}{c|}{\textbf{Aa}}& \multicolumn{1}{c|}{\textbf{A}} & \multicolumn{1}{c|}{\textbf{Baa}} & \multicolumn{1}{c|}{\textbf{Ba}} & \multicolumn{1}{c|}{\textbf{B}} & \multicolumn{1}{c|}{\textbf{Caa}} & \multicolumn{1}{c|}{\textbf{Ca}}& \multicolumn{1}{c|}{\textbf{C}}\\ \hline
\multicolumn{1}{|c|}{\textbf{Precision}} & 0.2 & 0.64 & 0.61 & 0.62 & 0.39 & 0.47 & 0.67 & 0.18 & 0.00
\\
\multicolumn{1}{|c|}{\textbf{Recall}}    & 0.07 & 0.30 & 0.61 & 0.64 & 0.53 & 0.56 & 0.33 & 0.28 & 0.00
  \\
\multicolumn{1}{|c|}{\textbf{Fscore}}    & 0.11 & 0.41 & 0.61 & 0.63 & 0.45 & 0.51 & 0.44 & 0.22 & 0.00
  \\ \hline
\end{tabular}
\caption{Precision, Recall and F$_{1}$-score for ANN with Embedding model}
\label{table:ANN_EMB_Results}

\end{table}
\FloatBarrier

\subsection{Models performances comparison}

\paragraph{Linear Regression}

The overall performance of Linear Regression model in term of Quadratic Weighted Kappa Metric is: $0.798 \pm 0.020$.
Details of Precision, Recall and F$_{1}$-score metrics are reported in \Cref{table:LinN_Results}.

\begin{table}[h!]
\centering

\begin{tabular}{c|r|r|r|r|r|r|r|r|r|}
\cline{2-10}
\textbf{}                                & \multicolumn{9}{c|}{\textit{Rating Classes}}                                                                                                                                                                             \\ \cline{2-10} 
\textit{}                                & \multicolumn{1}{c|}{\textbf{Aaa}} & \multicolumn{1}{c|}{\textbf{Aa}}& \multicolumn{1}{c|}{\textbf{A}} & \multicolumn{1}{c|}{\textbf{Baa}} & \multicolumn{1}{c|}{\textbf{Ba}} & \multicolumn{1}{c|}{\textbf{B}} & \multicolumn{1}{c|}{\textbf{Caa}} & \multicolumn{1}{c|}{\textbf{Ca}}& \multicolumn{1}{c|}{\textbf{C}}\\ \hline
\multicolumn{1}{|c|}{\textbf{Precision}} & 0.13 & 0.47 & 0.57 & 0.62 & 0.39 & 0.49 & 0.69 & 0.05 & 0.00
               \\
\multicolumn{1}{|c|}{\textbf{Recall}}    & 0.14 & 0.30 & 0.47 & 0.64 & 0.53 & 0.62 & 0.36 & 0.06 & 0.00
  \\
\multicolumn{1}{|c|}{\textbf{Fscore}}    & 0.13 & 0.37 & 0.51 & 0.63 & 0.45 & 0.55 & 0.48 & 0.05 & 0.00
  \\ \hline
\end{tabular}
\caption{Precision, Recall and F$_{1}$-score for Linear Regression model}
\label{table:LinN_Results}

\end{table}
\FloatBarrier

\paragraph{Logistic Regression}

For the Logistic Regression model (LR) all categorical features have been one-hot encoded. The overall performance of the model in term of Quadratic Weighted Kappa Metric is: $0.781 \pm 0.025$.
Details of Precision, Recall and F$_{1}$-score metrics are reported in \Cref{table:LR_Results}.

\begin{table}[h!]
\centering

\begin{tabular}{c|r|r|r|r|r|r|r|r|r|}
\cline{2-10}
\textbf{}                                & \multicolumn{9}{c|}{\textit{Rating Classes}}                                                                                                                                                                             \\ \cline{2-10} 
\textit{}                                & \multicolumn{1}{c|}{\textbf{Aaa}} & \multicolumn{1}{c|}{\textbf{Aa}}& \multicolumn{1}{c|}{\textbf{A}} & \multicolumn{1}{c|}{\textbf{Baa}} & \multicolumn{1}{c|}{\textbf{Ba}} & \multicolumn{1}{c|}{\textbf{B}} & \multicolumn{1}{c|}{\textbf{Caa}} & \multicolumn{1}{c|}{\textbf{Ca}}& \multicolumn{1}{c|}{\textbf{C}}\\ \hline
\multicolumn{1}{|c|}{\textbf{Precision}} & 0.63 & 0.66 & 0.69 & 0.67 & 0.58 & 0.57 & 0.63 & 0.42 & 0.00 \\
\multicolumn{1}{|c|}{\textbf{Recall}}    & 0.36 & 0.49 & 0.62 & 0.79 & 0.50 & 0.56 & 0.66 & 0.28 & 0.00  \\
\multicolumn{1}{|c|}{\textbf{Fscore}}    & 0.45 & 0.56 & 0.66 & 0.73 & 0.54 & 0.56 & 0.64 & 0.33 & 0.00  \\ \hline
\end{tabular}
\caption{Precision, Recall and F$_{1}$-score for Logistic Regression model}
\label{table:LR_Results}

\end{table}
\FloatBarrier

\paragraph{Artificial Neural Network without Embedding}

In order to point out the impact of embeddings on the overall model performance, the results of an Artificial Neural Network without Embedding (ANN) have been presented. The overall performance of the model in term of Quadratic Weighted Kappa Metric is: $0.861 \pm 0.009$.
Details of Precision, Recall and F$_{1}$-score metrics are reported in \Cref{table:ANN_Results}.

\begin{table}[h!]
\centering

\begin{tabular}{c|r|r|r|r|r|r|r|r|r|}
\cline{2-10}
\textbf{}                                & \multicolumn{9}{c|}{\textit{Rating Classes}}                                                                                                                                                                             \\ \cline{2-10} 
\textit{}                                & \multicolumn{1}{c|}{\textbf{Aaa}} & \multicolumn{1}{c|}{\textbf{Aa}}& \multicolumn{1}{c|}{\textbf{A}} & \multicolumn{1}{c|}{\textbf{Baa}} & \multicolumn{1}{c|}{\textbf{Ba}} & \multicolumn{1}{c|}{\textbf{B}} & \multicolumn{1}{c|}{\textbf{Caa}} & \multicolumn{1}{c|}{\textbf{Ca}}& \multicolumn{1}{c|}{\textbf{C}}\\ \hline
\multicolumn{1}{|c|}{\textbf{Precision}} & 0.35 & 0.55 & 0.66 & 0.74 & 0.55 & 0.58 & 0.74 & 0.19 & 0.50 \\
\multicolumn{1}{|c|}{\textbf{Recall}}    & 0.57 & 0.49 & 0.65 & 0.71 & 0.66 & 0.67 & 0.54 & 0.28 & 0.20  \\
\multicolumn{1}{|c|}{\textbf{Fscore}}    & 0.43 & 0.52 & 0.65 & 0.73 & 0.60 & 0.62 & 0.63 & 0.23 & 0.29  \\ \hline
\end{tabular}
\caption{Precision, Recall and F1-score for ANN without Embedding model}
\label{table:ANN_Results}

\end{table}
\FloatBarrier

\begin{figure}[h!]
     \centering
	\includegraphics[scale=0.8]{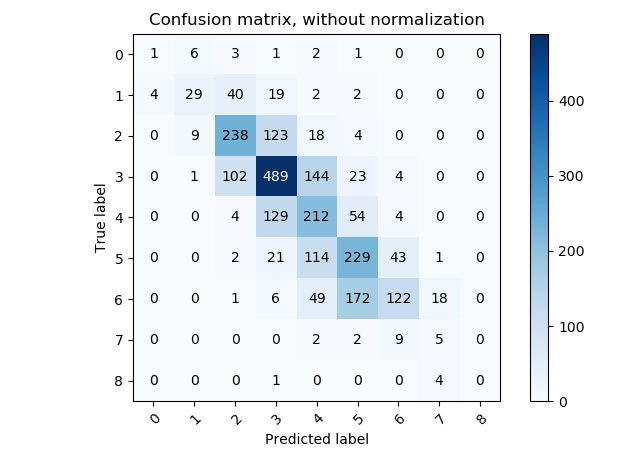}
     \caption{Absolute confusion matrix for ANN EMB model}
\end{figure}
\FloatBarrier

\begin{figure}[h!]
     \centering
	\includegraphics[scale=0.8]{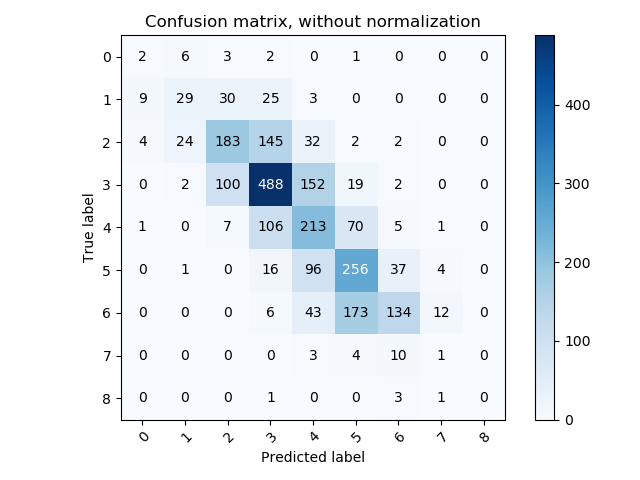}
     \caption{Absolute confusion matrix for Linear Regression model}
\end{figure}
\FloatBarrier

\begin{figure}[h!]
     \centering
	\includegraphics[scale=0.8]{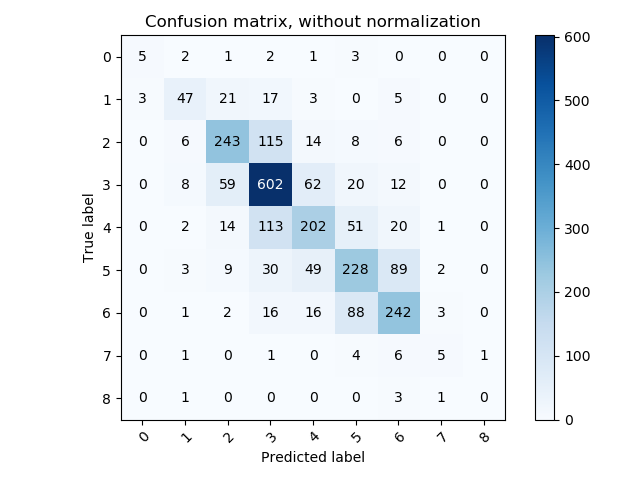}
     \caption{Absolute confusion matrix for Logistic Regression model}
\end{figure}
\FloatBarrier

\begin{figure}[h!]
     \centering
	\includegraphics[scale=0.8]{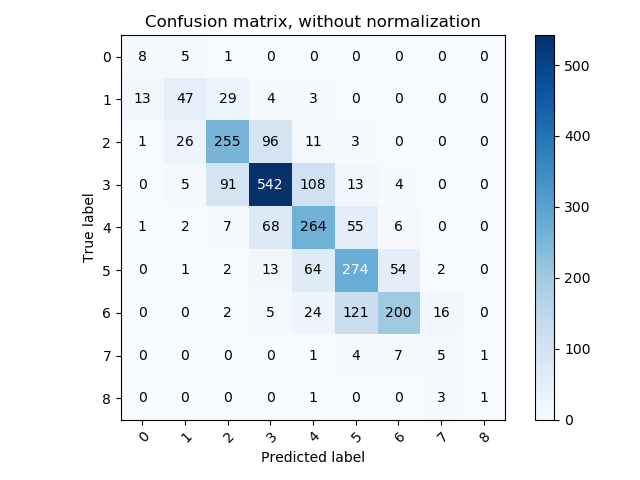}
     \caption{Absolute confusion matrix for ANN without Embedding model}
\end{figure}
\FloatBarrier

\subsection{Model explainability}

One main source of discussion about the use of neural networks regards its reputation as being a "\emph{black box}" in relation to its very nested, non-linear relationship between independent variables and final targets. Indeed, by employing complex machine learning models, more and more accuracy is often achieved on the detriment of results interpretability. \\Several methods have recently been developed to support users in interpreting the output of complex models. A novel unified approach to explain model outputs is \emph{SHAP} (\textsl{SH}apley \textsl{A}dditive ex\textsl{P}lanation) \cite{lundberg2017unified}, which leverages the idea of \emph{Shapley regression values} \footnote{Shapley regression value is defined as the average marginal contribution of a feature value over all possible combinations} to assign each feature an importance value for a particular prediction. SHAP values quantify the magnitude and direction (positive or negative) of a feature's effect on a prediction via an additive feature attribution method. In simple words, SHAP builds model explanations by asking, for each prediction $i$ and feature $j$, how $i$ changes when $j$ is removed from the model. Since SHAP considers all possible predictions for an instance using all possible combinations of feature inputs, it can guarantee both consistency and local accuracy. More in details, SHAP method computes Shapley values from \emph{coalitional game theory}. The feature values of a data instance act as players\footnote{note that a player can be an individual feature value or  a group of feature values} in a coalition: Shapley values suggest how to fairly distribute the “\emph{payout}” (i.e. the prediction) among the features. 

\paragraph{Force Plot}

\begin{figure}[h!]
     \centering
	\includegraphics[scale=0.65]{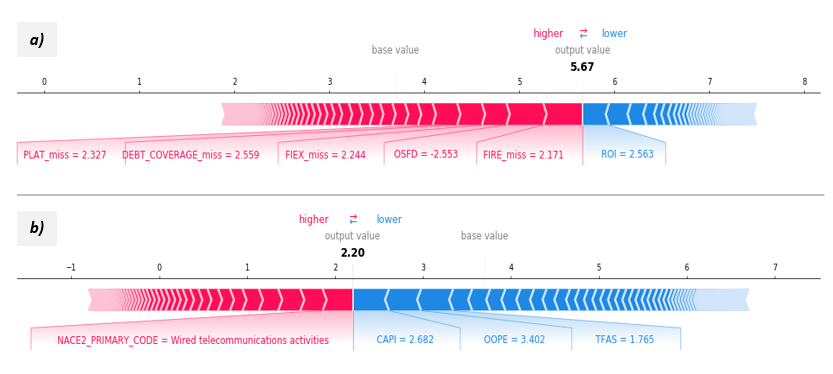}
     \caption{SHAP force plots for a sample of rating predictions. (a) The company rated in this example shows a prediction of 5.67 in the rating scale, corresponding to C rating class. The details of model's feature description are reported in Appendix \ref{appendix}. \\In particular, ROI equal to 2.563 (value after preprocessing) decreases its rating, while the absence of Financial Revenue (FIRE) feature increases its rating score. (b) This company has a rating score of 2.20; belonging to the telecommunications sector increases the company rating score, while its values of Capital (CAPI), Other Operating Expenses (OOPE) and Tangible Fixed Assets (TFAS) decrease the prediction value}
     \label{fig:force_plot}
\end{figure}

Each Shapley value can be visualized as a “\emph{force}” that either increases or decreases the prediction from its baseline, i.e. the average model output over the training dataset we passed. In \Cref{fig:force_plot} a sample of SHAP force plots are reported. Each Shapley value is an arrow that pushes to increase (positive value shown in red) or decrease (negative value shown in blue) the prediction. The balance of all these "forces" results to the actual model prediction.

\paragraph{SHAP feature importance}

\begin{figure}[h!]
     \centering
	\includegraphics[scale=0.4]{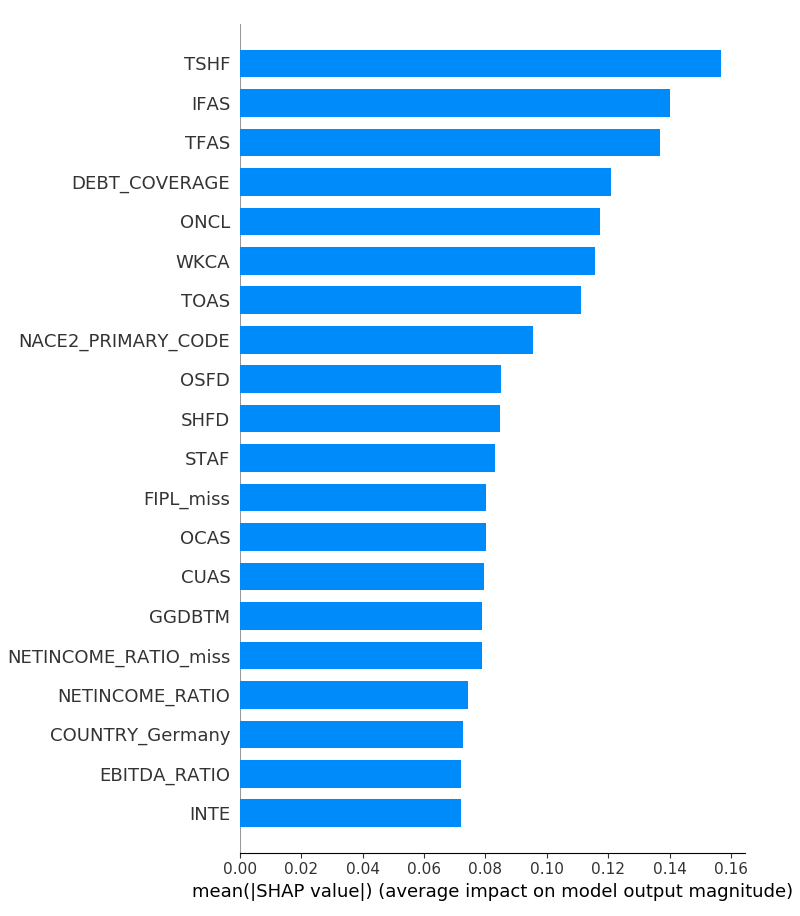}
     \caption{SHAP feature importance measured as the mean absolute Shapley values for ANN EMB model. The details of model's feature description are reported in Appendix \ref{appendix}}
     \label{fig: feature_importance}
\end{figure}

SHAP force plots described so far are explanations for individual predictions, but Shapley values can be combined into a global interpretation of the model output. A way to provide global explanation of the prediction is computing the average of the absolute Shapley values per feature across the dataset: features with large absolute Shapley values are "important" for model's output interpretation. This technique, known as \emph{SHAP feature importance}, is an alternative to \emph{permutation feature importance} which consists in measuring the deterioration of predicting performance of the model via random feature permutation. Whilst permutation feature importance is based on the decrease in model performance, SHAP feature importance is based on magnitude of feature attributions. In \Cref{fig: feature_importance} a set of features are plotted in term of decreasing feature importance.

\paragraph{SHAP summary plot}

\begin{figure}[h!]
     \centering
	\includegraphics[scale=0.4]{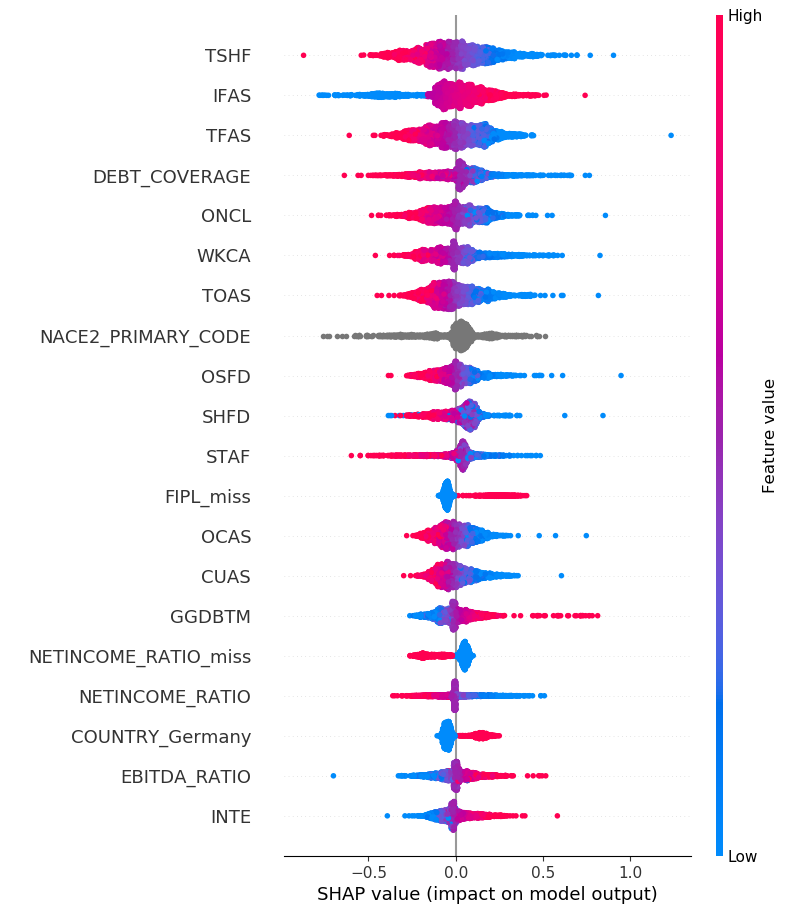}
     \caption{SHAP summary plot for ANN EMB model. Note that high values of TSHF reduce rating score. The details of model's feature description are reported in Appendix \ref{appendix}}
     \label{fig: summary_plot}
\end{figure}

The SHAP summary plot (as reported in \Cref{fig: summary_plot}) combines feature importance with feature effects. For each feature shown in the y-axis, and ordered according to their importance, each point on the plot represents the Shapley value (reported along the x-axis) for a given prediction. The color of each point represents the impact of the feature on model output from low (i.e. blue) to high (i.e. red). Overlapping points are littered in y-axis direction, so we get a sense of the distribution of the Shapley values per feature. 

\newpage
\FloatBarrier


\section{Conclusions}
\label{section:conclusions}

The aim of this work was to show that Neural Networks might represent a superior technique for calibrating and predicting ratings relative to other modelling approaches currently used in the banking sector. \\As shown by empirical results in the sections above, the model displays a better performance on investment grade ratings rather than on lower grades, although this might be explained by the relative low number of observations being used in the training. Further studies could be conducted in presence of a richer dataset, in order to provide more accuracy along the whole range of rating grades. \\Apart from performance considerations, however, we must point out that embedding of words included in \emph{Nace2 Description} allows for further generalization capabilities and flexibility of our model. In fact any description, even the ones never seen during the training phase, can be safely treated given the ability of the embeddings to correctly locate each word along the embedding space. Moreover, the embedding vectorization produces some interesting side-effects, such as the possibility of clustering of Nace2 categories in groups according to the values assigned to each dimension of the embedding vector, which then might be of help in sector analysis and visualization of correlations between neighbouring sectors. 

We concluded that, as ratings are used for numerous applications in financial markets and in banking regulation, and as rating agencies do not reveal the detail of their methodology for assigning ratings, it is crucial having a tool that allows the mimicking of the rating process. This means our proposal of Neural Network approach for shadow rating prediction has various fields of application, ranging from bank's risk management, to business decision process. In particular, ANN models could represent a useful tool in informing and supporting the analyst in the decision process.

\section*{Acknowledgements}
We would like to thank for the support our colleagues Lorenzo Giada, Andrea Riciputi and Davide Cuccio.
We are grateful to Corrado Passera for encouraging our research.

\bibliographystyle{ieeetr}
\bibliography{mybib} 

\newpage
\begin{appendices}
\section{Model's Features descriptions}\label{appendix}

\begin{table}[h!]
\centering
\resizebox{\textwidth}{!}{\begin{tabular}{lllll}

AV     & Added value                                                                             &  & OCLI      & Other current liabilities                      \\
CAPI   & Capital                                                                                 &  & OFAS      & Other fixed assets                             \\
CASH   & Cash \& cash equivalent                                                                 &  & ONCL      & Other non-current liabilities                  \\
CF     & Cash Flow                                                                               &  & OOPE      & Other operating expenses                       \\
COST   & Costs of goods sold                                                                     &  & OOPI      & Other operating items                          \\
CRED   & Creditors                                                                               &  & OPPL      & Operating P/L {[}=EBIT{]}                      \\
CREDR  & Credit rating, average                                                                  &  & OPRE      & Operating Revenue                              \\
CUAS   & Current assets                                                                          &  & OSFD      & Other shareholders funds                       \\
CULI   & Current liabilities                                                                     &  & PL        & P/L after tax                                  \\
DEBT   & Debtors                                                                                 &  & PLAT      & P/L for period {[}=Net income{]}               \\
DEPR   & Depreciation \& Amortization                                                            &  & PLBT      & P/L before tax                                 \\
DIV    & Stockmarket dividends index                                                             &  & PROV      & Provisions                                     \\
EBTA   & EBITDA                                                                                  &  & RCB       & Interest rate, central bank policy             \\
EMPL   & Number of employees                                                                     &  & RD        & Research \& Development expenses               \\
ENVA   & Enterprise value                                                                        &  & REONIA    & Interest rate, EONIA                           \\
EXEX   & Extr. and other expenses                                                                &  & RLG5      & Interest rate, 5-year Government Bond Yield    \\
EXPT   & Export revenue                                                                          &  & RSH6M     & Interest rate, 6-month                         \\
EXRE   & Extr. and other revenue                                                                 &  & RSWAP10YR & Interest rate, 10-year government bond yield   \\
EXTR   & Extr. and other P/L                                                                     &  & RSWAP10YR & Interest Rate Swap, 10-year                    \\
FIAS   & Fixed assets                                                                            &  & RSWAP2YR  & Interest Rate Swap, 2-year                     \\
FIEX   & Financial expenses                                                                      &  & RSWAP30YR & Interest Rate Swap, 30-year                    \\
FIPL   & Financial P/L                                                                           &  & RSWAP5YR  & Interest Rate Swap, 5-year                     \\
FIRE   & Financial revenue                                                                       &  & RTBILL3M  & 3 Month Treasury Bill Mid Yield (AVG, p.a.) \\
GDP    & GDP, real, LCU                                                                          &  & SHFD      & Shareholders funds                             \\
GGDBTM & \begin{tabular}[c]{@{}l@{}}Government debt,\\   Maastricht definition, LCU\end{tabular} &  & SMP       & Share price index, Global Equity Indices       \\
GROS   & Gross Profit                                                                            &  & STAF      & Costs of employees                             \\
IFAS   & Intangible fixed assets                                                                 &  & STOK      & Stock                                          \\
INTE   & Interest paid                                                                           &  & TAXA      & Taxation                                       \\
LOAN   & Loans                                                                                   &  & TFAS      & Tangible fixed assets                          \\
LTDB   & Long term debt                                                                          &  & TOAS      & Total assets                                   \\
MATE   & Material costs                                                                          &  & TSHF      & Total shareh. funds \& liab.                   \\
NCAS   & Net current assets                                                                      &  & TURN      & Sales                                          \\
NCLI   & Non-current liabilities                                                                 &  & WKCA      & Working Capital                                \\
OCAS   & Other current assets                                                                    &  &           &                                               
\end{tabular}}
\caption{Model Features Description}
\label{tab:caption}
\end{table}
\FloatBarrier

\end{appendices}

\end{document}